\documentclass[prx,two column,show pacs,show keys,superscript address,superscript reference,floatfix]{revtex4-1}
\usepackage[colorlinks=true,citecolor=blue,linkcolor=blue,urlcolor=red]{hyperref}

\usepackage[sort&compress]{natbib}
\usepackage{graphicx,times}
\usepackage{color}
\usepackage{sansmath}
\usepackage{dsfont}
\usepackage{amssymb}
\usepackage{amsthm}

\usepackage{subeqnarray}
\usepackage{bm}
\usepackage{diagbox}
\usepackage{makecell}
\usepackage{amsmath}

\usepackage{mathtools}
\DeclarePairedDelimiter\abs{\lvert}{\rvert}%

\begin{document}

\title{Enhanced connectivity of quantum hardware with digital-analog control}

\author{Asier Galicia} 
\affiliation{Department of Physical Chemistry, University of the Basque Country UPV/EHU, Apartado 644, 48080 Bilbao, Spain}

\author{Borja Ramon} 
\affiliation{Department of Physical Chemistry, University of the Basque Country UPV/EHU, Apartado 644, 48080 Bilbao, Spain}

\author{Enrique Solano} 
\affiliation{Department of Physical Chemistry, University of the Basque Country UPV/EHU, Apartado 644, 48080 Bilbao, Spain}
\affiliation{IKERBASQUE, Basque Foundation for Science, Maria Diaz de Haro 3, 48013 Bilbao, Spain}
\affiliation{International Center of Quantum Artificial Intelligence for Science and Technology (QuArtist) \\ and Department of Physics, Shanghai University, 200444 Shanghai, China}

\author{Mikel Sanz} \email{mikel.sanz@ehu.es} 
\affiliation{Department of Physical Chemistry, University of the Basque Country UPV/EHU, Apartado 644, 48080 Bilbao, Spain}

\begin{abstract}
Quantum computers based on superconducting circuits are experiencing a rapid development, aiming at outperforming classical computers in certain useful tasks in the near future. However, the currently available chip fabrication technologies limit the capability of gathering a large number of high-quality qubits in a single superconducting chip, a requirement for implementing quantum error correction. Furthermore, achieving high connectivity in a chip poses a formidable technological challenge. Here, we propose a hybrid digital-analog quantum algorithm that enhances the physical connectivity among qubits coupled by an arbitrary inhomogeneous nearest-neighbour Ising Hamiltonian and generates an arbitrary all-to-all Ising Hamiltonian only by employing single-qubit rotations. Additionally, we optimize the proposed algorithm in the number of analog blocks and in the time required for the simulation. These results take advantage of the natural evolution of the system by combining the flexibility of digital steps with the robustness of analog quantum computing, allowing us to improve the connectivity of the hardware and the efficiency of quantum algorithms.
\end{abstract}

 \keywords{}
\maketitle
\section{Introduction}

Quantum computation has emerged in recent years as a promising technology which
aims at solving problems such as the factorization of a composite number
\cite{Shor}, studying quantum field theories \cite{Martinez_2016,Klco_2018}, simulating quantum chemistry \cite{GALMSSL2016,KMTTBChG2017}, fluid dynamics \cite{MSLESS2015}, or the simulation of complex systems \cite{Lanyon_2011_DQS,Simulation1,Simulation2,SSSPS2016} more efficiently than classical computation. Another reason for the increasing
interest on the field of quantum computation is due to Grover's algorithm
\cite{Grover_1996}, which shows quadratic speedups when compared against
classical search algorithms. This technology can be implemented in different quantum
platforms, such us trapped ions, photonics, or superconducting qubits, among others. In this last platform, strong efforts have been performed to show an example in which a quantum processor outperforms any classical computer, milestone recently achieved by Google \cite{google_supremacy}. This flourishing technology still presents a considerable number of challenges to be solved, such as increasing the number of high-quality qubits in a single quantum processor or achieving interactions among all qubits. These problems characterize the so-called Noise Intermediate-Scale Quantum (NISQ) devices \cite{NISQ}, quantum chips comprising 50-100 qubits, which are still affected by significant noise.

With the objective of optimizing the currently available resources for quantum computation and, ultimately,
implementing a useful quantum computer in the NISQ era, an alternative paradigm of quantum computing called digital-analog quantum computation
(DAQC) \cite{DAQC,DAQC_theorica_cirucits} has recently been proposed. This approach aims at improving the
performance of digital quantum computers by taking advantage of the robustness of the natural dynamics of the chip, instead of turning off the internal
interactions \cite{DMNBT2002}. This approach has been proved theoretically viable when an
all-to-all (ATA) Ising Hamiltonian describes the natural dynamics of the quantum
processor, meaning that DAQC achieves universal computation using ATA analog
blocks along with single qubit rotations (SQR). It has also been given a
decomposition of the quantum Fourier transform (QFT) using the DAQC paradigm
\cite{DAQC-QFT}. Furthermore, simulations show an improvement in the QFT
implementation when reasonable assumptions regarding noise are made. At this point, it is remarkable the banged DAQC approach, in which the Hamiltonian is always on and the SQR are performed on top of the analog dynamics. It seems to pose advantages in the fidelity of the unitary matrices simulated when compared to the stepwise approach. However, the intrinsic interactions among the different qubits in a chip are not
necessarily well described by an ATA homogeneous Hamiltonian. In fact, a
physical quantum chip is expected to present only nearest-neighbour (NN)
interactions, since ATA connections require a prohibitively
increasing amount of wiring among the qubits.

In this Article, we design an
algorithm that optimally simulates an arbitrary ATA Ising Hamiltonian employing
as a resource a given inhomogeneous NN Ising model and SQR. This is achieved
employing $\mathcal{O}(3L^2)$ analog blocks, i.e. NN evolutions, with $L$ the number of qubits of the chip. Even though the particular dynamics considered
as a resource is the ZZ Ising Hamiltonian, the proposed algorithm could be
extended to other dynamics, such as the XX+YY Ising Hamiltonian. The simulation
of the Hamiltonian is optimal in the number of blocks and the simulation time
required.

\section{Graph representation of an Ising Hamiltonian}
The Ising Hamiltonian for $L$ qubits can be interpreted as a weighted graph of
$L$ vertices, where the weight of the edge connecting the vertex $i$ to the
vertex $j$ is $g_{ij}$. If two vertices $i$ and $j$ are not connected, $g_{ij}=
0$.

In this representation, an ATA Ising Hamiltonian of $L$ qubits becomes a
complete graph $K_L$, i.e. a graph with edges among every possible vertex
without repetition. On the other hand, the NN Ising Hamiltonian is represented
as a Hamiltonian path, that is, a path visiting all the possible vertices
only once.

\begin{figure}
	\centering \includegraphics[width=\linewidth]{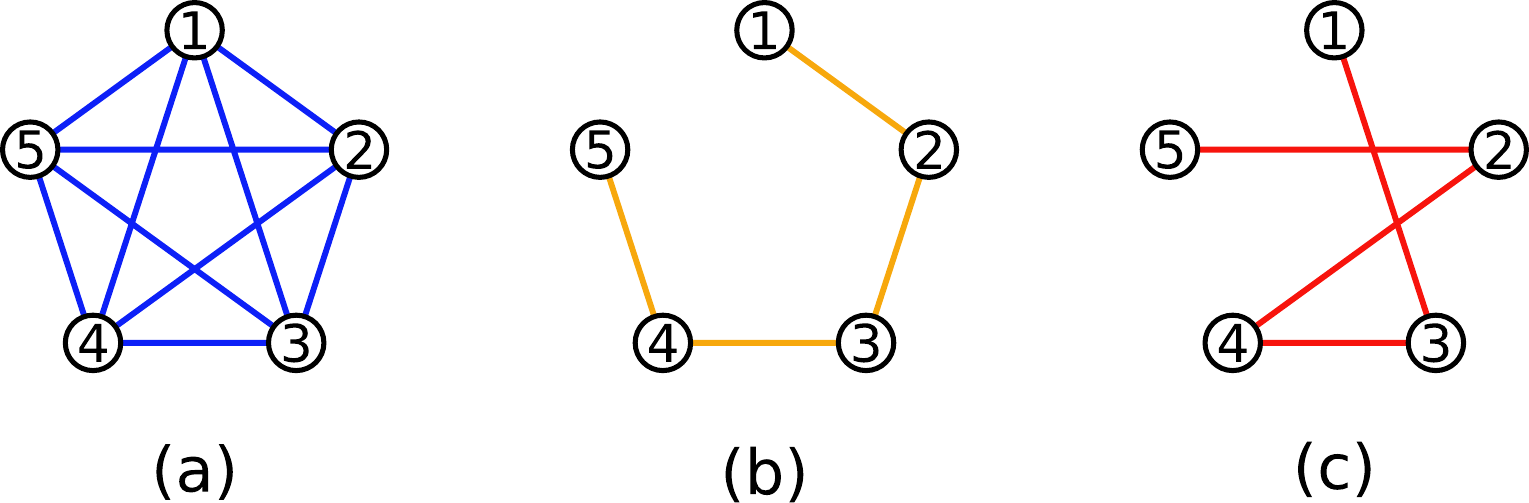}
	\caption{Examples of Ising Hamiltonians as their graph representation. Figure
    (a) shows a complete $K_5$ graph, which represents a $H_\text{ATA}(g)$ Hamiltonian
    for 5 qubits. Figure (b) shows a Hamilton path which represents a
    $H_\text{NN}(g)$ for 5 qubits. Figure (c) shows another Hamiltonian path with the
    vertex permutation $P = [1,3,4,2,5]$.}
	\label{fig:sec1:graphandpaths}
\end{figure}

It is noteworthy to mention that an arbitrary Hamiltonian path is represented by
a permutation of all the vertices in the graph. To recover a Hamiltonian path
from a given vertex permutation, it suffices to connect with an edge the
vertices that are adjacent in the permutation. An example of a complete graph,
together with two Hamiltonian paths, are represented in Fig.~\ref{fig:sec1:graphandpaths}, where we also show the vertex permutation of the
Hamiltonian paths.

Notice that we are currently dealing with ZZ interactions and thus, different Hamiltonian path evolutions commute. This means that the
final evolution will be the sum of all the Hamiltonian paths weighted by their
respective evolution time. This last statement is summarized in the
equation
\begin{equation}
  \label{eq:Behaviour_of_multiple_Hamiltonian_Paths.}
  \prod_ie^{i t_i HP_i(g_j)} = e^{i \sum_i t_i HP_i(g_j)},
\end{equation}
where $HP_i(g_j)$ is a Hamiltonian that describes a ZZ interaction which has a
graph representation of a Hamiltonian path with wights $g_j$. This is understood
in the graph representation as having as final graph the sum of all the weighted
Hamiltonian paths.

Our first task is then to split the complete graph, which represents the ATA
Hamiltonian, into a set of Hamiltonian paths that will be later simulated using
our resource (the NN Hamiltonian). This will allow us to efficiently decompose the ATA
evolution in terms of Hamiltonian paths.

\begin{figure}
  \centering \includegraphics[width=1\linewidth]{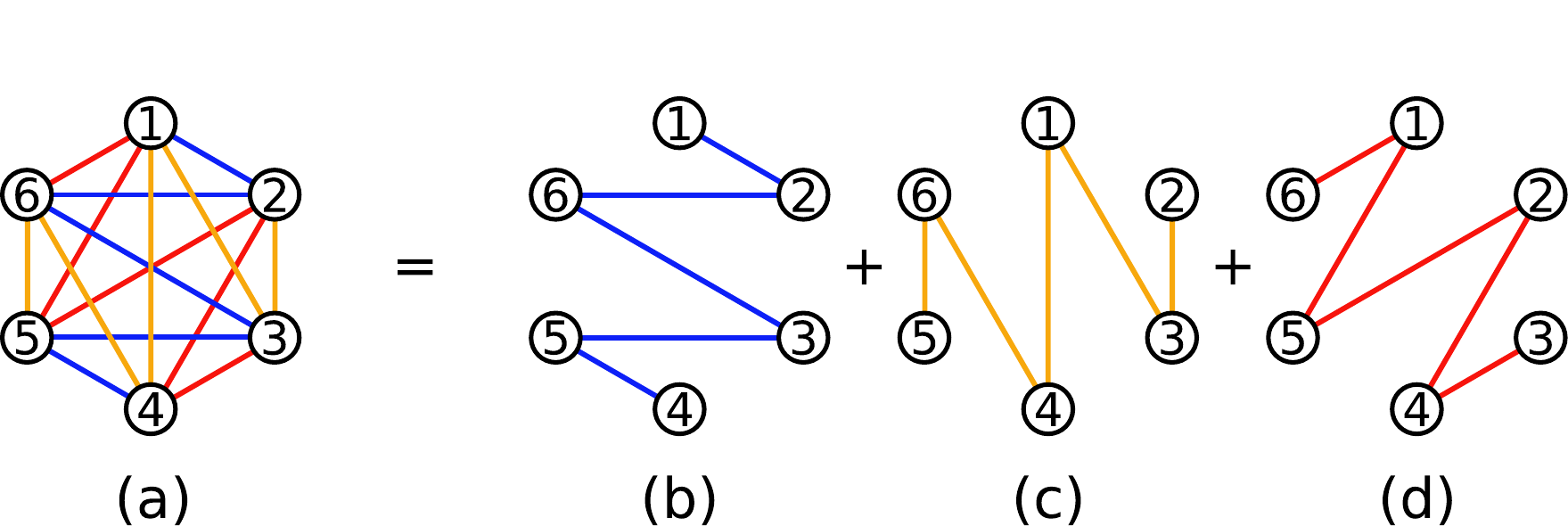}
  \caption{An example of how to fill exactly a complete graph of 6 vertices
    using Hamiltonian paths. Figure (a) represents the complete graph we are trying to
    get. Figures (b), (c) and (d) correspond to the Hamiltonian paths we use. Figure (b)
    is built by starting in the first node, going forward to the next node,
    then 2 nodes backward, 3 forward... The other two Hamiltonian paths are
    obtained rotating the first one. From this construction technique we obtain
    Hamiltonian paths defined by the permutations of Eq.
    \ref{eq:Decomposition_of_an_ATA}. The permutations that define (b), (c), and
    (d) are respectively $P_6^1 = [1,2,6,3,5,4]$ , $P_6^2 = [2,3,1,4,6,5]$ and
    $P_6^3 = [3,4,2,5,1,6]$.}
  \label{fig:ExampleDecomposition}
\end{figure}

Partitioning a complete graph into a set of Hamiltonian paths resembles the
Hamiltonian decomposition problem, which is about partitioning a complete graph
into a set of Hamiltonian cycles. This last problem was solved by
Walecki \cite{WaleckiHamiltonianPathDecomposition,WaleckiHamiltonianPathDecomposition2}
in 1890, who used a construction in which one Hamiltonian cycle is rotated to
get all the cycles that composes the complete graph. Using a similar
decomposition schematized for 6 qubits in Fig. \ref{fig:ExampleDecomposition},
we can decompose the complete graph into a set of disjoint Hamiltonian paths.
These Hamiltonian paths are characterized by the vertex permutation
\begin{equation}\label{eq:Decomposition_of_an_ATA}
  P_L^k(j)  =  \left\{
    \begin{array}{lcc}
      (k-1+\frac{j}{2})\text{ mod } L\;+1, &   \text{if}  & j \text{ even} \\
      \\ (k-1 - \frac{j-1}{2})\text{ mod } L \;+1 &  \text{if} & j \text{ odd}  \\
    \end{array}	\right.,
\end{equation}
where $P_L^k \in S_L$, being $S_L$ the symmetric group for $L$ elements,
$L$ the number of qubits and $k\in\mathbf{Z}$ such that $1\leq~k\leq~L/2$. $j$
represents the $j$-th position of the vertex permutation. Note that $k$ is just
a label for the Hamiltonian path.

It is also noteworthy to mention that this partition is only valid for an even number of
qubits. When dealing with an odd number of qubits, and hence, and odd number of
vertex in the graph representation, we use the notion of single perfect
matching. This involves using the same Hamiltonian path construction of
Eq. \ref{eq:Decomposition_of_an_ATA}, but allowing one Hamiltonian path to overlap
with the rest. This will pose no problem in our later construction, since we will
be able to selectively turn off the desired interactions.

In Fig. \ref{fig:ExampleDecomposition}, we show an example for 6 qubits where we
also depicted the corresponding Hamiltonian paths. Hence, the problem know
consists on efficiently simulating these Hamiltonian paths to obtain the ATA
evolution.

\section{Swapping gates}\label{sec:swapping-gate}

The next step is to obtain each of the Hamiltonian path connections using a NN Hamiltonian as
a resource. For that, we will change the connections using a SWAP-like gate $U$
that performs the operations
\begin{align}\label{eq:Swapping_gate:SwapOperationRequirements}
  U(\sigma_z\otimes I )U^\dagger  = I\otimes \sigma_z \nonumber,\\
  U(I\otimes \sigma_z)U^\dagger  = \sigma_z\otimes I .
\end{align} 

The gate that changes the action of an arbitrary operator in a qubit to another
qubit is the SWAP gate, defined as
\begin{equation}
  U_\text{SWAP} = e^{i\frac{\pi}{4}( X^1X^2 + Y^1Y^2 + Z^1Z^2) },
\end{equation} 
where the superindices $1$ and $2$ refer to the qubit on which the gate acts.
However, as we only need to change $Z$ gates, we have some degrees of freedom
available. More precisely, the most general unitary gate that fulfils Eq.
\ref{eq:Swapping_gate:SwapOperationRequirements} is
\begin{align}\label{eq:Chapter3:Udecom}
  U(\alpha,\beta,\gamma)= &R^1_z\left[\pi\left(\frac{\gamma -\alpha}{2}  + \frac{1}{2} + \beta\right)\right]\nonumber\\&e^{i\frac{\pi}{4}( X^1X^2 + Y^1Y^2 + (\gamma + \alpha) Z^1Z^2) }\nonumber\\& R^1_z\left[\pi\left(\frac{\gamma-\alpha}{2}  - \frac{1}{2} - \beta\right)\right],
\end{align}where $R^1_z[\theta] = e^{i\sigma^1_z\frac{\theta}{2}}$ and $\alpha,\beta,\gamma \in \mathbb{R} $.

Since we will later build this gate using inhomogeneous Ising Hamiltonians from
the DAQC perspective, we will set $\alpha = \gamma= 0$ and $\beta =
-\frac{1}{2}$, obtaining the so-called iSWAP gate. The choice of
parameters will minimize the amount of single qubit gates and analog blocks
required, since
\begin{align}\label{eq:Chapter3:iSWAP}
  U_\text{iSWAP}=  e^{i\frac{\pi}{4}( X^1X^2 + Y^1Y^2) }.
\end{align}
We will focus now on obtaining a sequence of iSWAP gates acting on adjacent
qubits that efficiently transforms a system with NN connections into a system
with the desired Hamiltonian path connections. The reason to restrict ourselves to adjacent
iSWAP gates is that we want to decompose them using NN Ising Hamiltonians as a resource.

For the sake of clarity, we will use $U_{ij}$ to represent the iSWAP gate
between qubits $i$ and $j$. We will first show how this operation transforms a
system with NN couplings. The result will be a system represented by a
Hamiltonian path in its graph representation.

If we sandwich $\sigma_z^k \sigma_z^l$, a gate that applies the $Z$ gate
to the qubits $k$ and $l$, with the gate $U_{ij}$, we obtain
\begin{equation}\label{eq:Chapter4:TransformationiSWAP}
  U_{ij} \sigma_z^k\sigma_z^l U_{ij}^\dagger = \sigma_z^{\tau_{ij}(k)} \sigma_z^{\tau_{ij}(l)},
\end{equation} 
where we have defined the function $\tau_{ij}$ as a permutation of the indices
$i$ and $j$, that is, a transposition. More precisely, if $k \neq i,j$,
$\tau_{ij}(k) = k$, otherwise, $\tau_{ij}(i) = j$ and $\tau_{ij}(j) = i$.
Basically, the $U_{ij}$ gate changes $\sigma_z$ gates acting on qubit $i$ to act
on qubit $j$.

Due to the unitarity of the iSWAP operation, its action on a NN Hamiltonian
evolution can be written as $U_{ij} e^{itH(g)_{NN}} U_{ij}^\dagger =
e^{itU_{ij}H(g)_{NN} U_{ij}^\dagger}$. Therefore, the final Hamiltonian results
in
\begin{align}\label{eq:una ecuacion}
  U_{ij}H_\text{NN}(g) U_{ij}^\dagger  &= \sum_{k=1}^{L-1} g U_{ij}\sigma_z^k\sigma_z^{k+1} U_{ij}^\dagger \nonumber
  \\&=\sum_{k=1}^{L-1} g \sigma_z^{\tau_{ij}(k)}\sigma_z^{\tau_{ij}(k+1)}.
\end{align}

The initial vertex permutation defining the system's NN coupling was $\text{P} =
[1,2,3,...,L]$. After the iSWAP operation, the permutation that defines our
system is $\text{P'} =
[\tau_{ij}(1),\tau_{ij}(2),\tau_{ij}(3),...,\tau_{ij}(L)]$. This approach is
straightforwardly generalized to a system with arbitrary connections. This means
that, after sandwiching with iSWAP gates a system with certain connections, it
interchanges the ones that had the two qubits being affected by the iSWAP gate.
In our case, since we will be dealing with systems represented by a Hamiltonian
path, the iSWAP operation reduces to a transposition in the corresponding vertex
permutation.

As the set of transpositions that our NN resource can implement,
$\{\tau_{i~i+1}\}$ for $i \in [1,L-1]$, is a generator of the symmetric group
$S_L$, we can obtain any desired permutation using the correct transpositions.
This means that we can simulate the evolution of any system with couplings
represented by a Hamiltonian path using as resource the system with NN
couplings. For example, the Ising Hamiltonian that represents Fig.
\ref{fig:sec1:graphandpaths}(c), which we will call $H_1$, is just obtained
using the following transformations $H_1(g) =
U_{34}U_{23}H_\text{NN}(g)U^\dagger_{23}U^\dagger_{34}$.

For the sake of simplicity, we will define a sequence of transpositions to be
\begin{equation}\label{eq:sec:efficient_iSWAP_decomposition}
  S_{i\rightarrow j}  = \left\{
    \begin{array}{ll}
      \tau_{i\;i+1} \tau_{i+2\;i+3}  \cdots \tau_{j-1\;j}& \text{if } i< j  \\ 
      \mathbb{I} & \text{if } i\ge j\end{array} \right. .
\end{equation} 

The permutations defined in Eq. \ref{eq:Decomposition_of_an_ATA} can be composed in
terms of two groups of sequences,
\begin{align}\label{eq:GroupofSeq}
  G_1(k) &=
           S^{2k-2}_{1 \rightarrow 2}  S^{2k-3}_{2 \rightarrow 3}  \cdots \nonumber\\ &S^4_{1 \rightarrow 2k-4}   S^3_{2 \rightarrow 2k-3}  S^2_{1 \rightarrow 2k-2}   S^1_{2 \rightarrow 2k-1} \\\label{eq:GroupofSeq2}
  G_2(k,L) &=
             S^{L-2k-1}_{L-1 \rightarrow L}  S^{L-2k-2}_{L-2 \rightarrow L-1}   \cdots \nonumber \\ &S^4_{2k+4\rightarrow L-1}   S^3_{2k+3 \rightarrow L}    S^2_{2k +2 \rightarrow L-1}   S^1_{2k +1 \rightarrow L},
\end{align}
where $k$ and $L$ refer to the permutation $P_L^k$ we are building and the
sequences have been labeled to make clear the order of application. That is,
$P_L^k = G_1(k)G_2(k,L)$. In Appendix \ref{Appsec:Group_demonstration}, we proof
that these are indeed the groups of sequences needed to obtain the desired
permutations of Eq. (\ref{eq:Decomposition_of_an_ATA}). Note that both groups of
sequences commute between them.

\begin{figure*}
	\centering
  \includegraphics[width=\textwidth]{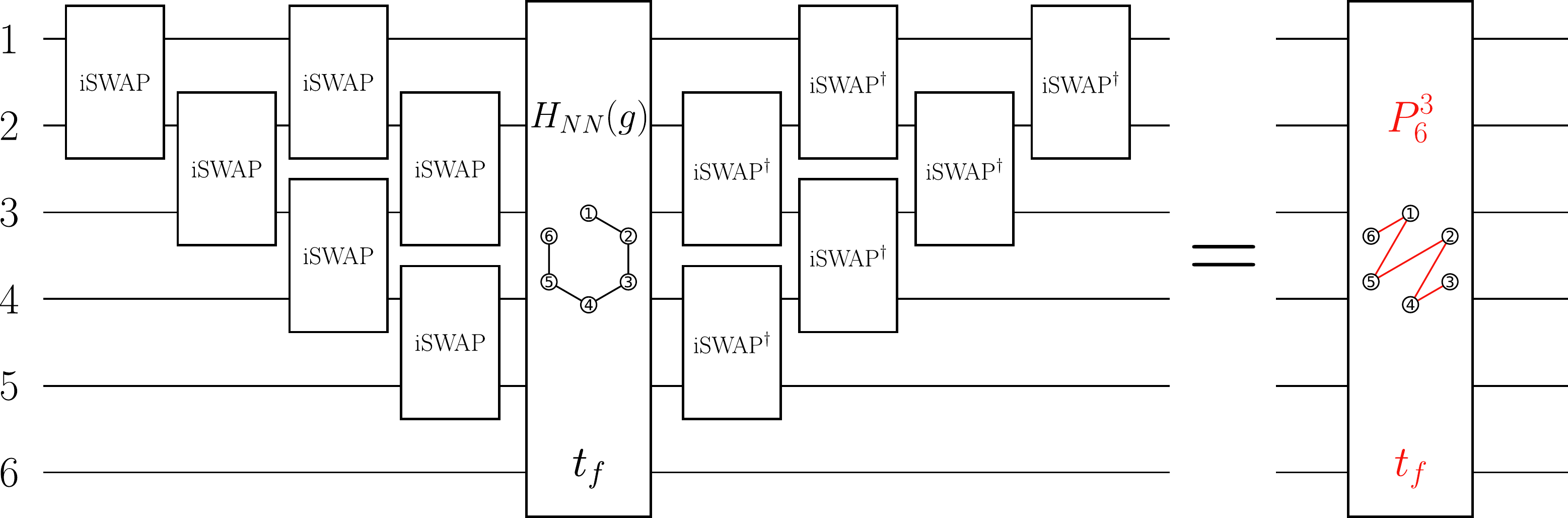}
	\caption{Circuit that simulates the evolution of a Hamiltonian
    path with vertex permutation $P^3_6$. Each of the surrounding columns of
    iSWAP gates is related with one of the sequences belonging to the groups
    defined in Eq. \ref{eq:GroupofSeq}.}
	\label{fig:vertexandedges}
\end{figure*}

Let us now show an example for the case of 6 qubits. In order to obtain the
permutation $P^3_6$ from Fig. \ref{fig:vertexandedges}, we need to apply the
transpositions defined in Eq. \ref{eq:GroupofSeq}, which are $G_1(3)$ and
$G_2(3,6)$. However, in the particular case of $G_2(3,6)$, since $2\times k +1 =
7$ and $L=6$, $G_2(3,6) = \mathbb{I}$. The sequences applied are $G_1(3) = S^4_{
  1 \rightarrow 2} S^3_{ 2 \rightarrow 3 } S^2_{ 1 \rightarrow 4 } S^1_{
  2\rightarrow 5 }$. The circuit that implements this set of transpositions
using iSWAP gates is shown in Fig. \ref{fig:vertexandedges}, where each column of iSWAP gates represents one sequence of transposition.

To sum up, in this section we have shown an algorithm that, using adjacent iSWAP
gates and the NN Ising Hamiltonian as resource, is able to simulate the
evolution of an ATA Hamiltonian for the case of an even number of qubits. Let us now simplify the circuit so that it requires the smallest possible amount of
iSWAP gates.

\section{Simplification of the circuit}
In this section, we will show an optimized version of the previously discussed
circuit. Besides, we will also give a decomposition of the circuit in terms of
ZZ gates, which will be useful later.

\begin{figure*}
	\centering
  \includegraphics[width=1\linewidth]{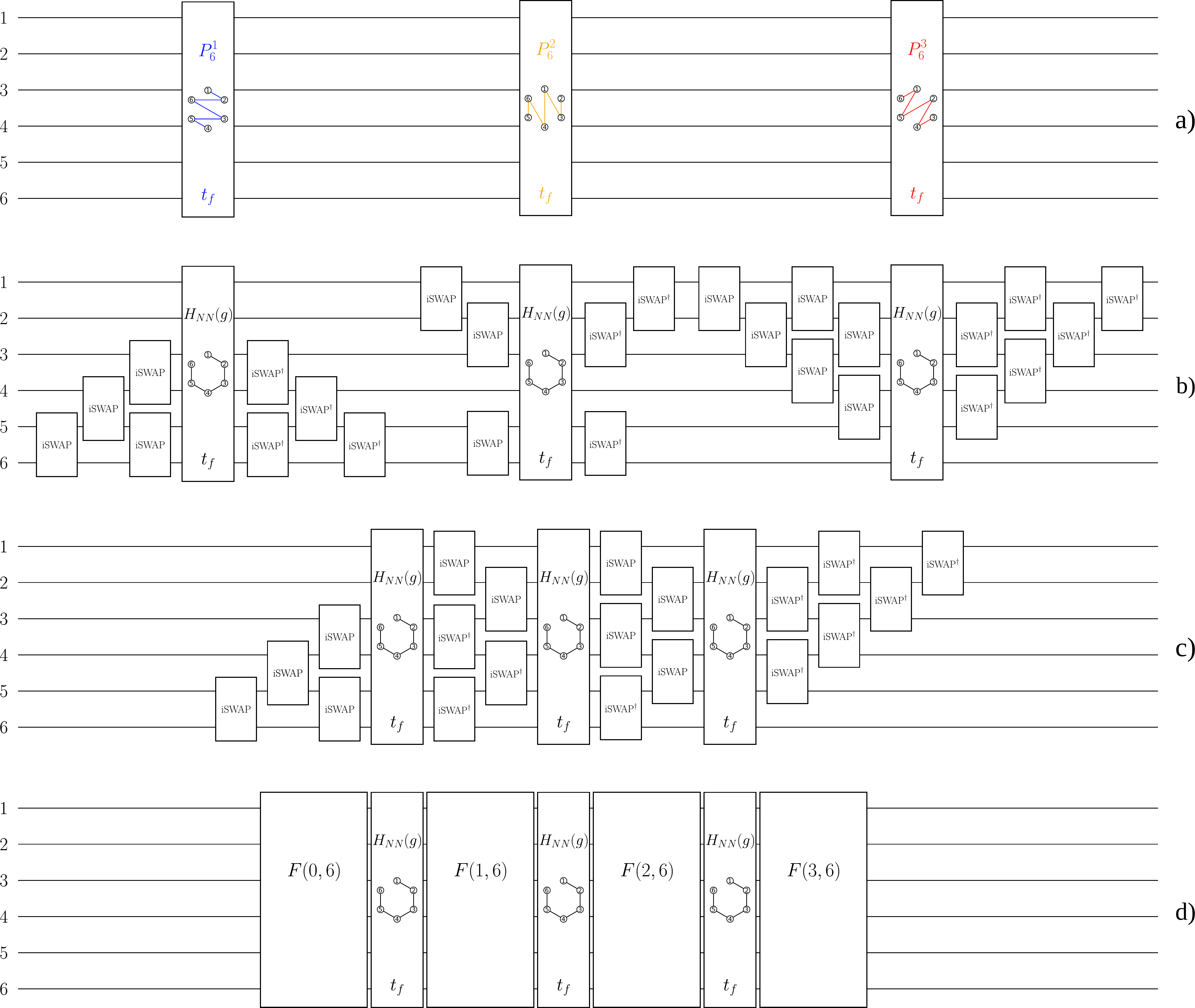}
	\caption{Simplified circuit simulating an ATA Hamiltonian evolution. Figure a) shows the three
    Hamiltonian paths that form the ATA evolution. Figure b) is obtained by
    introducing the iSWAP gates that transform a NN Hamiltonian in the desired
    Hamiltonian paths. This gates are described by the group of transpositions
    defined in Eq. \ref{eq:GroupofSeq}. The set of transpositions obtained with
    this equation is translated into a set of iSWAP/$\text{iSWAP}^{\dagger}$
    gates surrounding the NN Hamiltonian as discussed in Section
    \ref{sec:swapping-gate}. In c) we simplify the circuit by making use of
    $\text{iSWAP }\text{iSWAP}^\dagger = \mathbb{I}$. These are the set of iSWAP
    gates defined in Eq. \ref{eq:F_gates}, which lead to the circuit depicted in
    d).}
	\label{fig:FullCircuitDigital}
\end{figure*}

Since we now need to combine the set of Hamiltonian paths $\{P^k_L\}$ to obtain
the desired ATA evolution, it is possible to simplify the total number of iSWAP
($U$) gates needed. The total circuit is described by the set of gates
\begin{align}\label{eq:F_gates}
  F(k,L) = \prod_{i=2, i \text{ even}}^{2k+1} U_{i,i+1}&\prod_{i=2k +1,i \text{ even}}^{L-1} U^\dagger_{i,i+1}\nonumber
  \\\nonumber\prod_{i=1, i \text{ odd}}^{2k} U_{i,i+1}\prod_{i=2k+1,i \text{ odd}}^{L-1}& U^\dagger_{i,i+1},\quad k \in[1,\frac{L}{2}-1]\\
  F\left(0,L\right)= \prod_{j = \frac{L}{2}}^{j=1} \left[ \prod_{i=2j-1, i \text{ odd}}^{L-1}\right.&\left. U_{i,i+1}\prod_{i=2j-2,i \text{ even}}^{L-2} U_{i,i+1} \right],\nonumber
  \\
  F\left(\frac{L}{2},L\right)  =\prod_{j = 1}^{j=\frac{L}{2}}\left[ \prod_{i=1, i \text{ odd}}^{L-2j} \right.&\left. U^{\dagger}_{i,i+1}\prod_{i=2,i \text{ even}}^{L+1-2j} U^{\dagger}_{i,i+1} \right]
                                                                                                               .
\end{align}
The final ATA evolution will be described as
\begin{equation}\label{eq:OptimizedCircuit}
	e^{it_fH_\text{ATA}} = \prod_{k = 1}^{\frac{L}{2}}F(k,L)e^{it_fH_\text{NN}}F(k,L)F(0,L).
\end{equation}
In Appendix \ref{AppSec:DesmonstrationOptimizedCircuit}, we show the
demonstration leading to this simplified circuit. In Fig.
\ref{fig:FullCircuitDigital}, we show an example of the simplification for the case
of 6 qubits.

\begin{figure*}
	\centering
  \includegraphics[width=1\linewidth]{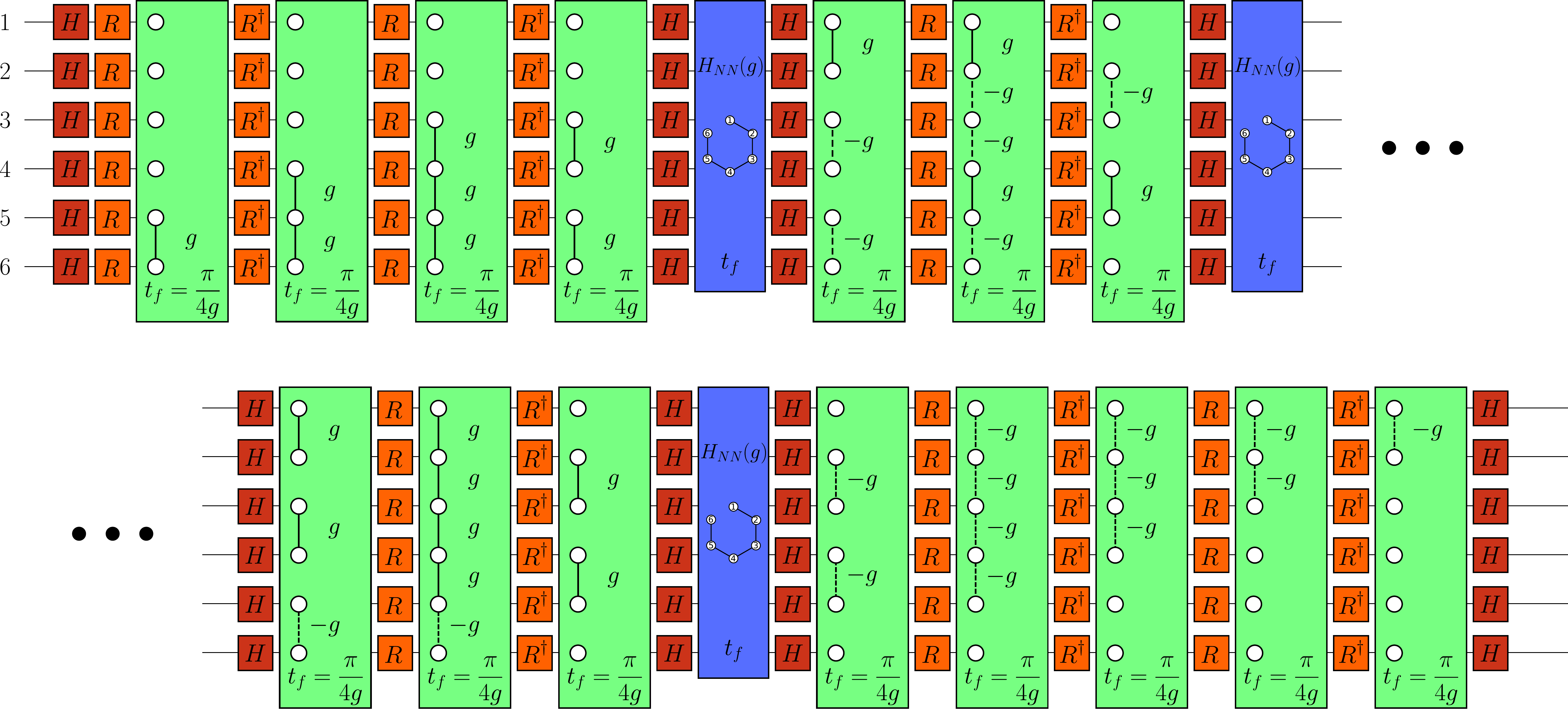}
	\caption{This circuit is obtained by replacing each iSWAP in Fig.
    \ref{fig:FullCircuitDigital} by the expression shown in Eq.
    \ref{eq:Chapter3:iSWAP}, where each of the exponentials have been multiplied
    by the appropriated single qubit gates to transform them into ZZ interactions.
    Then, these interactions are gathered and expressed as evolutions of NN
    Ising Hamiltonians, characterized by the different couplings $g$ and their
    evolution time $t_f$. Each of this inhomogeneous Hamiltonian evolutions can
    be implemented using the DAQC circuit discussed in Section
    \ref{sec:simulating-an-arbitrary-inhomogeneous-hamiltonian}. The SQR employed in this circuit are the Hadamard gate ($H$) and the
    gate $R$, defined as $R = HSH$ where $S$ is the phase gate.}
	\label{fig:FullCircuit}
\end{figure*}

Using Eq. \ref{eq:Chapter3:Udecom}, we can decompose the gates $F(k,L)$ into
a set of SQR and $ZZ_{i,j}(\phi = g_jt_f)$ gates acting
on adjacent qubits. Grouping parallel $ZZ_{ij}$ gates naturally leads to a
circuit which can be implemented using the DAQC protocol, as explained
in the following section. An example for the case of 6 qubits is shown in Fig.
\ref{fig:FullCircuit}.

We now take into account that $k$ consecutive parallel set of iSWAP gates can
be decomposed into a $k+1$ analog blocks plus some SQR. Since
the total number of $F(k,L)$ for $k\in [1,\frac{L}{2}-1]$ gates is
$\frac{L}{2}-1$ and each one requires two parallel sets of iSWAP gates, we require
a total of $\frac{3L}{2}-3$ analog blocks to generate these gates. We also need
$3 \left(\frac{L}{2}-2\right)$ analog gates to implement the $F(0,L)$ set of
gates and $3 \left(\frac{L}{2}-1\right)$ analog gates to implement $F\left(
  \frac{L}{2} , L \right)$. Moreover, we need $\frac{L}{2}$ analog blocks more
evolving during a time $t_f$. The total amount of inhomogeneous analog blocks
needed is then $5L-12$.

In the following section, given a NN inhomogeneous
Hamiltonian, we will derive an algorithm to simulate the evolution of an arbitrary NN inhomogeneous Hamiltonian efficiently, both in time and in the number of analog
blocks. This is the
last step before obtaining an algorithm to simulate an arbitrary inhomogeneous ATA
Hamiltonian.

\section{Simulating an arbitrary inhomogeneous Hamiltonian}\label{sec:simulating-an-arbitrary-inhomogeneous-hamiltonian}
In this section, we will show an algorithm to simulate the evolution of an
arbitrary inhomogeneous NN Hamiltonian under the paradigm of DAQC, employing a
similar algorithm to the one shown in Ref.~\cite{DAQC}. In this case, our resource
will be a fixed inhomogeneous NN Hamiltonian. The algorithm we use has the
following three advantages over the one shown in Ref.~\cite{DAQC}: (i) It works
for an arbitrary number of qubits; (ii) It requires the minimum amount of analog
blocks; (iii) It optimizes the time required for the simulation.

\begin{figure}
	\centering \includegraphics[width=1\linewidth]{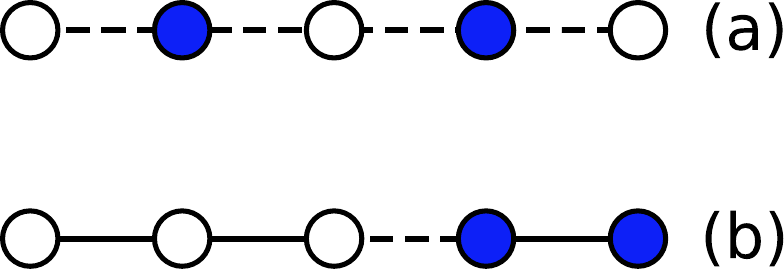}
	\caption{Coloured graphs representing the surrounding of X gates in the
    evolution of a NN system for 5 qubits. A coloured node corresponds to a
    qubit surrounded by X gates. The dotted lines represent the change of sing
    in the coupling. Fig. (a) represents the inversion of all couplings, which
    is the same as inverting the time evolution of the system. Fig. (b)
    represents the inversion of only one of the couplings.}
	\label{fig:coloredGraph}
\end{figure}

We will first need to notice that it is possible to selectively change the sign
of any desired combination of couplings. This is done by surrounding some of the
qubits with $X$ gates. We represent the action of the $X$ gates by colouring the
corresponding qubits in the graph representation (See Fig.
\ref{fig:coloredGraph}). In order to change the sign of the desired combination
of couplings, it suffices to colour differently the qubits connected to the
desired couplings. In Appendix \ref{AppSec:ColorGraphs} we prove that this can
be done for a NN chain with an arbitrary length. In Fig. \ref{fig:coloredGraph}, we
change the sign of all the couplings in (a) and the sign of just one coupling in
(b).

We will now decompose a $H_\text{NN}(g'_j)$ evolution during a time $t_f$ into a set
of $H_\text{NN}(g_j)$ evolutions that have been evolving during a time $t_n$ each,

\begin{align}
  &t_fH_\text{NN}(g'_j)= t_f\sum_{j=1}^{L-1}g'_j\sigma_z^{j} \sigma_z^{j+1} \nonumber\\=&
                                                                                     \sum_{j=1}^{L-1}\sum_{n=1}^{L-1} t_n g_j (-1)^{f_n(j)+f_n(j+1)}\sigma_z^{j} \sigma_z^{j+1} \nonumber \\=&
                                                                                                                                                                                               \sum_{j=1}^{L-1}\sum_{n=1}^{L-1} t_n \left(\prod_{k=1}^{L-1} X_k^{f_n(k)}\right)g_j \sigma_z^{j} \sigma_z^{j+1} \left(\prod_{k=1}^{L-1} X_k^{f_n(k)}\right)\nonumber \\=&
                                                                                                                                                                                                                                                                                                                                                                         \sum_{n=1}^{L-1} t_n \left(\prod_{k=1}^{L-1} X_k^{f_n(k)}\right)H_\text{NN}(g_j) \left(\prod_{k=1}^{L-1} X_k^{f_n(k)}\right),
\end{align}
where $X_i^j$ is an $X$ gate applied on the $i$-th qubit $j$-th times and
$f_n(k)$ is a binary function that determines whether an $X$ gate is being
applied in the $k$-th qubit during the $n$-th analog block, yet to be
determined. In the last step, we make use
\begin{equation}\label{eq:B-TRelation}
  b_j=\frac{g'_j}{g_j} = \sum_{n=1}^{L-1} M_{nj} \frac{t_n}{t_f},
\end{equation} where $M_{nj} = (-1)^{f_n(j) + f_n(j+1)}$. This defines the following system of linear equations,
\begin{equation}\label{eq:linearEquation}
  \textbf{b}=M\frac{\textbf{t}}{t_f}.
\end{equation}The matrix $M$ has only $\pm1$ entries. The interpretation of $M_{nj} =-1$ is that during the $n$-th analog block, the $j$-th coupling changes the sign. From now on, we will only focus on the $M$ matrix instead of the $f_n(j)$ functions. 

We will now assume without loss of generality that the following conditions hold $\forall j$
\begin{align}\label{eq:Constraint1}
	b_j&\ge b_{j+1}, \\
	\label{eq:Constraint2}
	\abs{b_1} &\ge \abs{b_j},\\
	\label{eq:Constraint3}
	b_j&>0.
\end{align} Without loss of generality, $b_j$ can always be relabelled and changed of sign to hold these inequalities.

Under these conditions, we propose the following $M$ matrix,
\begin{align}\label{eq:Mmatrix}
  M = \left[\begin{array}{ccccccccc}
              1  & 1   &  1 &  1   &\cdots&  1   &  1   & 1    & 1     \\ 
              -1  & 1   &  1 &  1   &\cdots&  1   &  1   & 1    & 1     \\ 
              -1  &-1   &  1 &  1   &      &  1   &  1   & 1    & 1     \\ 
              -1  &-1   & -1 &  1   &\ddots&  1   &  1   & 1    & 1     \\ 
              \vdots &\vdots&\vdots&\ddots&\ddots&\ddots&  \vdots    &\vdots& \vdots\\ 
              -1  &  -1   &  -1 &  -1   &\ddots&  1   &  1   &  1   &  1      \\ 
              -1  &  -1   &  -1 &  -1   &\ddots& -1   &  1   &  1   &  1      \\ 
              -1  &  -1   &  -1 &  -1   &\cdots& -1   & -1   &  1   &  1      \\ 
              -1  &  -1   &  -1 &  -1   &\cdots& -1   & -1   & -1   &  1      \\ 
            \end{array}  \right].
\end{align} 

After inverting this matrix (see Appendix \ref{AppSec:MatrixInversion}), Eq.
(\ref{eq:linearEquation}) leads to the time intervals
\begin{align}\label{eq:timeSol}
  \frac{t_k}{t_f}&=\frac{b_k-b_{k+1}}{2},  \\\label{eq:timeSol2}
  \frac{t_{L-1}}{t_f}&= \frac{b_1 + b_{L-1}}{2},
\end{align} with $k\in[1,L-2]\nonumber$. Recall that $t_k=0$ means that we do not need the $k$-th analog block for the simulation. 

Let us now prove that these solutions for $t_n$ imply the minimum
amount of analog blocks and that the time required for the simulation is
minimal. As, through Eqs. \ref{eq:timeSol} and \ref{eq:timeSol2}, we are mapping the
set of times $\{t_n\}_{n=1}^{L-1}$ to the values $\{b_j\}_{j=1}^{L-1}$, we need
at least as much different $t_n$ as $b_j$. Indeed, suppose that $b_j = b_{j'}$.
We can relabel them such that $j' = j+1$. Then, from Eq. \ref{eq:timeSol}, we get
that $t_j = 0$, so the total number of analog blocks is reduced by 1. Another
particularly relevant case is when $b_j = 0$ for $k$ different values, for which
the number of analog blocks needed is reduced by $k-1$.

The time required to simulate the desired $H_\text{NN}(g'_j)$ is defined as $t_\text{sim}
= \sum_{n=1}^{L-1} \abs{t_n}$. Note that we do not take into account the time
required to implement the $X$ gates since we are supposing that they are ideal
digital blocks, instantaneous. However, we believe that the
  circuit will still be minimum in time as long as we can parallelize the
  application of these gates. In Appendix \ref{AppSec:TimesSimultationMinimum},
we prove that, under the constrains of Eqs. \ref{eq:Constraint1},
\ref{eq:Constraint2} and \ref{eq:Constraint3}, $\min(t_{\text{sim}}) \equiv
t_\text{min} = \abs{b_1}t_f$. We also prove in Appendix
\ref{AppSec:TimesSimultationMinimum} that our circuit requires a time $t_\text{min}$
to perform the simulation of an arbitrary inhomogenous Hamiltonian, being this
the minimum time possible.

\begin{figure*}
	\centering
  \includegraphics[width=1\linewidth]{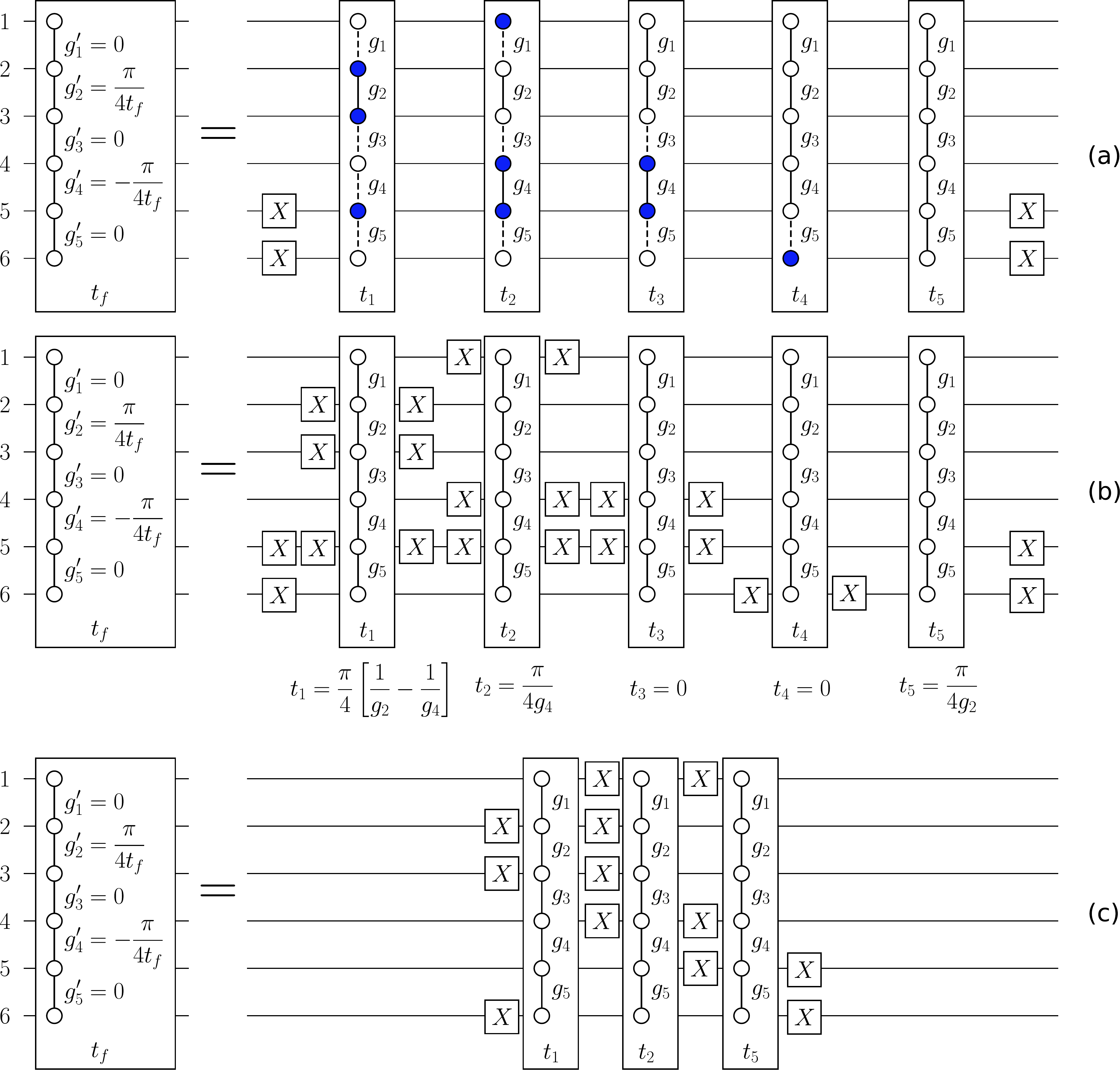}
	\caption{Implementation of the circuit discussed in
    Section \ref{sec:simulating-an-arbitrary-inhomogeneous-hamiltonian}. The
    digital block implemented is used in the circuit of Fig.
    \ref{fig:FullCircuit} to generate an iSWAP gate between qubits 2 and 3, in
    parallel with an $\text{iSWAP}^\dagger$ gate between qubits 4 and 5. The
    analog blocks shown in the RHS of figure (a) represent the evolution of a NN
    system, where the sign of some of the couplings have been inverted according
    to Eq. (\ref{eq:Mmatrix}). The different evolution times are determined by
    Eq. (\ref{eq:timeSol}) and (\ref{eq:timeSol2}). In order to
    meet the constraints imposed by Eq.
    (\ref{eq:Constraint1}-\ref{eq:Constraint3}), the coefficients $b_1$, $b_2$,
    $b_3$, $b_4$ and $b_5$ are equal to
    $\frac{g'_2}{g_2}$,$\frac{g'_4}{g_4}$,$\frac{g'_1}{g_1}$,$\frac{g'_3}{g_3}$
    and $\frac{g'_5}{g_5}$, respectively (we made the assumption that $g_4>g_2$
    and hence $\abs{b_2}>\abs{b_4}$). Since $b_4<0$, we need to change $g_4$ of
    sign in all the analog blocks, which is achieved by applying a $X$-gate in
    Fig. (a). The dashed lines in the analog block represent a coupling changed
    of sign. These dashed lines connects two qubits with different colour,
    whereas a solid line connects two qubits with the same colour. The change of
    sign of the $i$-th coupling is given by the $i$-th column of the matrix
    defined in Eq. (\ref{eq:Mmatrix}). For example, the first column of the
    matrix, shows that all the signs of $b_j$ belonging to the first block must
    be inverted except from $b_1$, which is related to $g_2$. Consequently, all
    the signs of the couplings must be inverted, except from $g_2$. In figure
    (b) we sandwiched each analog block with $X$ gates in the qubits that where
    coloured, representing in that way the same evolution of figure (a). Lastly,
    in figure (c) we simplified the previous circuit both by eliminating all the
    analog blocks with zero time evolution, and all unnecessary $X$ gates,
    taking into account that $XX = \mathbb{I}$.}
	\label{fig:Example_Inhomog_Homog_Circuit}
\end{figure*}

As an example, in Fig. \ref{fig:Example_Inhomog_Homog_Circuit} we represent the
implementation of one of the analog blocks shown in Fig. \ref{fig:FullCircuit}
e), required for a set of iSWAP gates. More precisely, the depicted block is
necessary for an iSWAP gate between the qubits 2 and 3 and an
$\text{iSWAP}^{\dagger}$ gate between the qubits 4 and 5.

It is noteworthy to mention that we require at least $L-1$ analog blocks to simulate the
evolution of an arbitrary inhomogeneous NN Hamiltonian.

Until now, we have shown an algorithm that simulates the evolution
of an homogeneous ATA Hamiltonian using as resource an inhomogeneous NN
Hamiltonian. Furthermore, it does so with $\mathcal{O}(5L^2)$ analog blocks,
since we need $\mathcal{O}(5L)$ inhomogeneous analog blocks which can be
simulated using $\mathcal{O}(L)$ analog blocks coming from the resource for
each. 

It is straightforward to modify the circuit in order to simulate an
inhomogeneous ATA Hamiltonian with negligible impact on the performance. It
suffices to change the blue blocks in the circuit of the Fig.
\ref{fig:FullCircuit} for an inhomogeneous Hamiltonian.

In order to implement this circuit for an odd number of qubits, $L$, we can use
the same set of Hamiltonian paths, $P^k_L$, of Eq.
\ref{eq:Decomposition_of_an_ATA}. In this case $k\in[1,\frac{L+1}{2}]$ and, in
order to obtain an homogeneous ATA Hamiltonian, we need to set to zero some of
the couplings used for the Hamiltonian path evolution representing
$P^{\frac{L+1}{2}}_L$. It should be noted that the number of analog blocks will
still $\mathcal{O}(3L^2)$. Even though we do not discuss here how to obtain the
iSWAP gates for this case. Similar techniques to those discussed in Appendix \ref{Appsec:Group_demonstration} can be employed.

\section{Conclusions}

We have shown that, within the DAQC paradigm, naturally arising evolutions can
be utilized to simulate any inhomogeneous ATA Ising Hamiltonian along with
SQR. In particular, we have designed an algorithm based on a
NN Ising Hamiltonian with $\mathcal{O}(5L^2)$ analog blocks,
where $L$ is the number of qubits in the chip. For this, we also discussed both
a digital approach that simulates an ATA system while having NN-like connections
and an algorithm that simulates under the DAQC paradigm the evolution of an
inhomogeneous Hamiltonian. This last algorithm has been proven to be efficient
in the number of analog blocks and in the simulation time required, as long as
we treat SQR as ideal gates. This protocol can be extended to
platforms described by different Hamiltonians, such us the XX+YY NN Ising
Hamiltonian.

\section*{Acknowledgements}
The authors acknowledge  support from the projects QMiCS (820505) and OpenSuperQ (820363) of the EU
  Flagship on Quantum Technologies, Spanish MINECO/FEDER FIS2015-69983-P, Basque Government IT986-16 and EU FET Open
  Grant Quromorphic. This material is also based upon work supported by
  the U.S. Department of Energy, Office of Science, Office of Advance Scientific
  Computing Research (ASCR), Quantum Algorithms Teams project under field work
  proposal ERKJ333.

\appendix

\section{\uppercase{Group demonstration}}
\label{Appsec:Group_demonstration}

In this appendix we proof that the combination of sequences $G_1(k)$ and
$G_2(k,L)$, defined in \eqref{eq:GroupofSeq}, decompose $P^k_L$ into a set of
adjacent transposition, that is, $P^k_L = G_1(k) G_2(k,L)$. For that, we will
first briefly discuss how to state the problem in terms of the matrix
representation of the permutation group \cite{SimetricGroupTheory}. For the sake
of clarity, we will denote by $T_k$ the matrix representation of a transposition
$\tau_k$.

The problem we are solving can be stated as obtaining a finite set of
transpositions $\{T_k\}_{k=1}^M$ that transforms the vector $\bold{b}$ with
components $b_i$ into the vector $\bold{b'}$ with components $b_{P^k_L(i)}$.
That is, $\prod_{k=1}^{M} T_k\bold{b} = \bold{b'}$. However, since $T^{-1} = T$,
this set of transpositions will hold that $\prod_{k=M}^1 T_k\bold{b'} =
\bold{b}$. The decomposition of $P_k^L$ is then $P^k_L = \tau^1\circ \tau^2
\circ \cdots \circ \tau^k $, where we use $b \circ a = b(a)$ to denote the order
of the operations.

The only restriction we will impose in the available transpositions is that they
need to be adjacent. Denoting by $T(i,j)$ the transposition $\tau_{i,j}$ that
transposes the elements $i$ and $j$, we note that $T(i,j)\bold{b}$ transposes
the elements of $\bold{b}$ in the positions $i$ and $j$. Hence, we can use
algorithms, such as the Bubble Short algorithm or its parallelized version, to
directly obtain an optimized set of transpositions $T^k$ that shorts a given
vector $\bold{b'}$. Indeed, the set of transpositions $G_1$ and $G_2$ have been
obtained using those techniques, though we considered necessary to give a closed
formula which we now prove to be correct.

We will now define $c_{i,j}$ as the operation that fulfills,
\begin{eqnarray}\label{eq:AppendixB:definingCij}
  c_{ik}(P_k^L(j)) = \left\{ \begin{array}{lccc}
                               P_L^k(j)&   \text{if}  & j &\neq i,k \\
                               P_L^k(i)&   \text{if}  & j &= k\\
                               P_L^k(k)&   \text{if}  & j &= i
                             \end{array}	\right..
\end{eqnarray} 
That is, it changes the position of the entry i with the entry k. We will define
$g_1$ and $g_2$ as $G_1$ and $G_2$ in Eq. but with all the $\tau_{ij}$
operations replaced by $c_{ij}$. From the representation we see that replacing
all the transpositions, $\tau_{ij}$ for $c_{ij}$, makes the new group of
operations $g_1$ and $g_2$ to fulfill the equation
\begin{equation}
  \label{eq:AppendixB:equationtoProve}
  g_1\circ g_2 \circ P_{L}^k = 1
\end{equation} 
where 1 stands for the identity permutation. This again resembles to shorting
and array defined by $P_L^k$ where the operations $c_{ij}$ are the changes in
positions made to that array. We continue by realizing that
\begin{equation}\label{eq:Permutations2}
  P^k_L(j)=\left\{
    \begin{array}{lcl}
      P^k_{2k}(j)		&   \text{if}  & j\leq 2k,  \\
      P^{L-2k}_{L-2k}(j-2k) + 2k  &  \text{if}   & j >  2k ,  \\
    \end{array}	\right.
\end{equation} 
that is, we obtain two commuting permutations. This is why two commuting groups
of sequences, $G_1(k)$ and $G_2(k,L)$, arise. Note that the second permutation,
which can be regarded as $P^{L'}_{L'}$ with $L' = L-2k$, is obtained from
\begin{align} &G_2(L') =S^{L'-1}_{L'-1 \rightarrow L'}\circ S^{L'-2}_{L'-2
    \rightarrow L'-1} \circ\cdots\nonumber\\&\cdots\circ S^4_{4\rightarrow L'-1}
  \circ S^3_{3 \rightarrow L'}\circ S^2_{2 \rightarrow L'-1} \circ S^1_{1
    \rightarrow L'} .
\end{align}
$G_2(L')$ differs from $G_2(k,L)$ by a factor of $2k$ that appears in all the
sequences. This extra factor in $G_2(k,L)$ comes from \eqref{eq:Permutations2},
where it appears adding to the permutation $P^{L-2k}_{L-2k}$. It has the effect
of changing the action of all transpositions from $\tau_{ij}$ to $\tau_{i+2k \;
  j+2k}$.

We will only prove how to short the permutation that fulfills $g_1(k) \circ
P^k_{2k} = 1 $ because proving that $g'_2(L') \circ P^{L'}_{L'}= 1$ requires the
same steps. We will prove this by induction.

Since, $g_1(1)$ is the identity operation and $P^1_{2} $ is the identity
permutation, it is clear that $g_1(1) \circ P^1_{2} = 1 $. We now suppose that
$g_1(k-1) \circ P^{k-1}_{2k-2} =1 $ and we prove that, with this condition, $
g_1(k) \circ P^{k}_{2k} = 1 $. Since $g_1(k) \circ P^{k}_{2k} = g_1(k-1) \circ
s_{1 \rightarrow 2k-2}\circ s_{2 \rightarrow 2k-1} \circ P^{k}_{2k} $, it
suffices to prove that $ s_{1 \rightarrow 2k-2}\circ s_{2 \rightarrow 2k-1}
\circ P^{k}_{2k}=P^{k-1}_{2k-2}$, where $s_{i\rightarrow j}$ is defined in Eq.
\ref{eq:sec:efficient_iSWAP_decomposition} but with all the $\tau_{ij}$
operations replaced by $c_{ij}$ operations. Notice that $s_{2 \rightarrow 2k-1}$
has the effect of changing the position of all the entries in $P^k_L$ except for
the last and the first one. All the numbers in an odd position change to their
left position and all the number in a even position change to their right
position. Hence, the permutation obtained from $\pi_1 = s_{2 \rightarrow 2k-1}
\circ P^{k}_{2k} $ results in
\begin{align*}
  \pi_1(j)  &= \left\{
              \begin{array}{lcl}
                P^k_{2k}(j+1)  	 	&   \text{if}  & j \text{ even } \\
                P^k_{2k}(j-1)        &  \text{if}   & j \text{ odd }  \\
                P^k_{2k}(1) 	   	&   \text{if}  & j = 1\\
                P^k_{2k}(2k)                   &  \text{if}   & j = 2k  \\
              \end{array}\right.\\
            &= \left\{
              \begin{array}{lcl}
                P^k_{2k}(j+1)  	 	&   \text{if}  & j \text{ even } \\
                P^k_{2k}(j-1)        &  \text{if}   & j \text{ odd }  \\
                2k                   &  \text{if}   & j = 2k  \\
              \end{array}\right.
\end{align*}
where we used $P^k_{2k}(0) = P^k_{2k}(1)$.

We now compute the operation $\pi_2 = s_{1 \rightarrow 2k-2}\circ \pi_1 $, which
changes the position of all the entries except from the last two. Hence,
\begin{align*}
  \pi_2(j)  &= \left\{
              \begin{array}{lcl}
                \pi_1(j-1)  	 	&   \text{if}  & j \text{ even } \\
                \pi_1(j+1)      &  \text{if}   & j \text{ odd }  \\
                \pi_1(2k-1)	   	&   \text{if}  & j = 2k-1\\
                \pi_1(2k)	                 &  \text{if}   & j = 2k  \\
              \end{array}\right.\\
            &= \left\{
              \begin{array}{lcl}
                P^k_{2k}(j-2) 	 	&   \text{if}  & j \text{ even } \\
                P^k_{2k}(j+2)      &  \text{if}   & j \text{ odd }  \\
                2k-1	   	&   \text{if}  & j = 2k-1\\
                2k	                 &  \text{if}   & j = 2k  \\
              \end{array}\right.\\
            &= \left\{
              \begin{array}{lcl}
                P^{k-1}_{2k-2}(j) 	 	&   \text{if}  & j<2k-1 \\
                2k-1	   	&   \text{if}  & j = 2k-1\\
                2k	                 &  \text{if}   & j = 2k  \\
              \end{array}\right.
\end{align*}
Since $g_1(k-1)$ does not affect to the positions $2k$ and $2k-1$,
\begin{align*}
  g_1(k-1)\circ \pi_1(j)  &= \left\{
                            \begin{array}{lcl}
                              g_1(k-1)\circ P^{k-1}_{2k-2}(j) 	 	&   \text{if}  & j<2k-1 \\
                              2k-1	   	&   \text{if}  & j = 2k-1\\
                              2k	                 &  \text{if}   & j = 2k  \\
                            \end{array}\right.\\
                          &= 1.
\end{align*}This completes the proof.

\section{\uppercase{Colored graphs demonstration}}
\label{AppSec:ColorGraphs}

In order to prove that we can selectively change the sign of a coupling in an
arbitrary length NN chain we will use an induction process. If $L$ is the number
of nodes in a NN chain, then, $k = L-1$ is the number of couplings.

Inverting the couplings for the $k=1$ case is trivial. Supposing that we can
selectively change the coupling sing of the $k = k'-1$, we will prove that we
can do it for the case $k = k'$. The construction relies in the fact that, in
order to change the sign of the new coupling, it suffices to change the colour
of the newly added node. The colour of the new node must be different form its
neighbour colour, which can always be achieved in the NN case. If we want the
sign of the $k'$ coupling not to change, we just need to colour the new node as
its neighbour.

\section{\uppercase{Demonstration of EQ.} \ref{eq:F_gates} }
\label{AppSec:DesmonstrationOptimizedCircuit}
In this appendix, we show how to obtain the gates defined in Eq.
\ref{eq:F_gates}. For that, we first define various set of gates that will
simplify the notation. We define $\tilde{S}$ to be the set of
$\text{iSWAP}^{\dagger}$ gates that are obtained from substituting $\tau_{ij}
\rightarrow \text{iSWAP}^{\dagger}_{ij}$ in Eq.
\ref{eq:sec:efficient_iSWAP_decomposition}. At the same time, $\tilde{G}$ is
defined to be the set of $\text{iSWAP}^{\dagger}$ gates that are obtained from
substituting $S\rightarrow \tilde{S}$ in Eq. \ref{eq:GroupofSeq}. Hence, it
holds that
\begin{equation}
  \label{eq:Appendix:HP-HNNrelation}
  \text{HP}(P^k_L) = \tilde{G_1}^{\dagger} \tilde{G_2}^{\dagger} H_\text{NN} \tilde{G_2} \tilde{G_1},
\end{equation} 
where $H_\text{NN}$ is the NN Hamiltonian and $\text{HP}(P^k_L)$ is the ZZ
interaction Hamiltonian described by the Hamiltonian Path $P^k_L$.

$F(k,L)$ describes the gates between the NN Hamiltonians that will be used to
implement the Hamiltonian paths with vertex permutation $P^k_{L}$ and
$P^{k=1}_{L}$. Hence
\begin{align}
  \label{eq:appendix:Fdefinition}
  \nonumber
 & F(k,L) =\nonumber \\ & \left\{
    \begin{array}{ll}
      \tilde{G}^{\dagger}_1(k+1)\tilde{G}_2^{\dagger}(k+1,L)\tilde{G}_1(k)\tilde{G}_2(k,L)& k \in [1,\frac{L}{2}-1],\\  
      \tilde{G}^{\dagger}_2(1,L)& k  = 0, \\
      \tilde{G}_1(\frac{L}{2})& k  = \frac{L}{2},
    \end{array}
  \right. 
\end{align}
where it is straight forward to prove that $F(k,L)$ has the form described in
Eq .\ref{eq:F_gates} for $k=0$ and $k=\frac{L}{2}$. For $ k \in
[1,\frac{L}{2}-1]$, we will prove that the set of iSWAP gates can be further
simplified to obtain the Eq. \ref{eq:F_gates}.

We first note that the following equations hold from the definition of
$\tilde{G}$,
\begin{align}
  \label{eq:Appendix:G_Relations}
  \tilde{G}_1(k+1) &=  \tilde{G}_1(k) \tilde{S}_{1\rightarrow 2k} \tilde{S}_{1\rightarrow 2k+1} \\
  \tilde{G}_2(k,L) &= \tilde{G}_2(k+1,L) \tilde{S}_{2k+2\rightarrow L-1}\tilde{S}_{2k+1\rightarrow L}.
\end{align}

Hence, it follows that
\begin{align}
  \label{eq:Appendix:Proof}
  \nonumber
  &\tilde{G}^{\dagger}_1(k+1)\tilde{G}_2^{\dagger}(k+1,L)\tilde{G}_1(k)\tilde{G}_2(k,L)\\
  \nonumber
  =&\tilde{S}^{\dagger}_{1\rightarrow 2k+1} \tilde{S}^{\dagger}_{1\rightarrow 2k}
  \tilde{G}^{\dagger}_1(k) \tilde{G}_2^{\dagger}(k+1,L)
  \tilde{G}_1(k)\tilde{G}_2(k,L)\\
  \nonumber
  =&\tilde{S}^{\dagger}_{1\rightarrow 2k+1} \tilde{S}^{\dagger}_{1\rightarrow 2k}
  \tilde{G}_2^{\dagger}(k+1,L)
  \tilde{G}_2(k,L)\\
  \nonumber
  =&\tilde{S}^{\dagger}_{1\rightarrow 2k+1} \tilde{S}^{\dagger}_{1\rightarrow 2k}
  \tilde{G}_2^{\dagger}(k+1,L)
  \tilde{G}_2(k+1,L) \tilde{S}_{2k+2\rightarrow L-1}\tilde{S}_{2k+1\rightarrow L}\\
  \nonumber
  =&\tilde{S}^{\dagger}_{1\rightarrow 2k+1} \tilde{S}^{\dagger}_{1\rightarrow 2k}
  \tilde{S}_{2k+2\rightarrow L-1}\tilde{S}_{2k+1\rightarrow L}.
\end{align}
where we used that $[\tilde{G}^{\dagger}_1(k), \tilde{G}_2^{\dagger}(k+1,L)] = 0
$ and $ \tilde{G}^{\dagger}_1(k)\tilde{G}_1(k)=
\tilde{G}_2^{\dagger}(k+1,L)\tilde{G}_2(k+1,L) = 0$. 

We now have that
\begin{equation}
\label{eq:appendix:Fdefinition2}
\nonumber
F(k,L) = \left\{
\begin{array}{ll}
\tilde{S}^{\dagger}_{1\rightarrow 2k+1} \tilde{S}^{\dagger}_{1\rightarrow 2k}
\tilde{S}_{2k+2\rightarrow L-1}\tilde{S}_{2k+1\rightarrow L}
& k \in [1,\frac{L}{2}-1],\\  
\tilde{G}^{\dagger}_2(1,L)& k  = 0, \\
\tilde{G}_1(\frac{L}{2})& k  = \frac{L}{2},
\end{array}
\right. 
\end{equation}
which when expressed in terms of the iSWAP gates, turns out to be the Eq. \ref{eq:F_gates}. 

\section{\uppercase{Inversion of matrix $M$}}
\label{AppSec:MatrixInversion}

For the inversion of the matrix $M$ of Eq. \ref{eq:Mmatrix} it suffices to make
row operations that transforms the $M$ into the identity matrix while the
identity matrix is transformed into $M^{-1}$. The required row operations are
\begin{eqnarray}
  r_i &=& \frac{r_i+r_1}{2} \qquad  i > 1,\\
  r_i &=& r_i-r_{i+1}\qquad i\in[1,L-2],
\end{eqnarray}where $r_i$ refers to the $i$-th row, $L$ is the number of qubits and hence $L-1$ is the dimension of $M$.

\section{\uppercase{Minimum time of simulation for arbitrary inhomogeneous NN
    Hamiltonian}}
\label{AppSec:TimesSimultationMinimum}

In this appendix we will prove that the solutions obtained in Eq.
\ref{eq:timeSol} give, under the constraints of Eq. \ref{eq:Constraint1},
\ref{eq:Constraint2} and \ref{eq:Constraint3}, the minimum value possible to
$t_\text{sim}$. This function is defined as
\begin{equation}
  t_\text{sim} = \sum_{n=1}^{L-1} \abs{t_n}.
\end{equation}

We will first prove that $\min(t_\text{sim}) = \abs{b_1 t_f}$. For that,
recall that Eq. \ref{eq:B-TRelation} defines the relation between $b_j$ and
$t_n$. Computing its absolute value, we obtain
\begin{align}
  \abs{b_j} &= \abs{\sum_{n=1}^{L-1} M_{nj} \frac{t_n}{t_f}} \le \sum_{n=1}^{L-1} \abs{\frac{t_n}{t_f}}\nonumber\\
            &\le \frac{t_\text{sim}}{t_f}.\nonumber
\end{align}Using the constraints, we know that $b_1$ is the maximum value of $b_j$ $\forall j$. This proves that   $\min
(t_\text{sim}) = \abs{b_1 t_f}$.

The solutions obtained in Eq. \ref{eq:timeSol} hold that $t_k >0 $ $\forall k$.
We now compute $\sum_{k=1}^{k=L-1}\abs{t_k} = \sum_{k=1}^{k=L-1}t_k = b_1 t_f$,
which proves that this set of solutions makes $t_\text{sim}$ minimum.

Notice that, even thought the constraints may seem too restrictive, $b_j$ can
always be relabelled or changed its sign in order to hold them.

\end{document}